\RequirePackage{tikz}
\documentclass[pdflatex, sn-standardnature]{sn-jnl}

\usepackage{amsthm,amsmath}

\usepackage{pdfpages}
\usepackage{enumerate}
\usepackage{tabularx}
\usepackage{lineno}
\usepackage{makecell}
\usepackage{bbm}

\usepackage{amsmath,graphicx,centernot}

\jyear{2024}%

\theoremstyle{thmstyleone}%
\theoremstyle{thmstyletwo}%

\usepackage{bm} 
\usepackage{breqn} 
\usepackage{systeme}  
\usepackage{amssymb} 
\DeclareMathOperator{\E}{E}
\DeclareMathOperator{\expit}{expit}
\DeclareMathOperator{\logit}{logit}

\usepackage{multirow}

\theoremstyle{thmstylethree}%

\usetikzlibrary{shapes,decorations,arrows,calc,arrows.meta,fit,positioning}
\tikzset{
    -Latex,auto,node distance =1 cm and 1 cm,semithick,
    state/.style ={ellipse, draw, minimum width = 0.7 cm},
    point/.style = {circle, draw, inner sep=0.04cm,fill,node contents={}},
    bidirected/.style={Latex-Latex,dashed},
    el/.style = {inner sep=2pt, align=left, sloped}
}
\tikzstyle{arrow} = [thick,->,>=stealth]

\begin{document}

\title[MedZIsc (Magics)]{A Statistical Framework for Co-Mediators of Zero-Inflated Single-Cell RNA-Seq Data}


\author*[1]{\fnm{Seungjun} \sur{Ahn}}\email{seungjun.ahn@mountsinai.org}
\author[2]{\fnm{Li} \sur{Chen}}\email{li.chen1@ufl.edu}
\author[3]{\fnm{Maaike} \sur{van Gerwen}}\email{maaike.vangerwen@mountsinai.org}
\author[4,5,6]{\fnm{Panos} \sur{Roussos}}\email{panagiotis.roussos@mssm.edu}
\author[2]{\fnm{Zhigang} \sur{Li}}\email{zhigang.li@ufl.edu}

\affil[1]{\orgdiv{Department of Population Health Science and Policy}, \orgname{Icahn School of Medicine at Mount Sinai}, \city{New York}, \state{NY}, \country{U.S.A}}
\affil[2]{\orgdiv{Department of Biostatistics}, \orgname{University of Florida}, \city{Gainesville}, \state{FL}, \country{U.S.A}}
\affil[3]{\orgdiv{Department of Otolaryngology - Head and Neck Surgery}, \orgname{Icahn School of Medicine at Mount Sinai}, \city{New York}, \state{NY}, \country{U.S.A}}
\affil[4]{\orgdiv{Center for Disease Neurogenomics}, \orgname{Icahn School of Medicine at Mount Sinai}, \city{New York}, \state{NY}, \country{U.S.A}}
\affil[5]{\orgdiv{Department of Psychiatry}, \orgname{Icahn School of Medicine at Mount Sinai}, \city{New York}, \state{NY}, \country{U.S.A}}
\affil[6]{\orgdiv{Department of Genetics and Genomic Sciences}, \orgname{Icahn School of Medicine at Mount Sinai}, \city{New York}, \state{NY}, \country{U.S.A}}


\abstract{Single-cell RNA sequencing (scRNA-seq) has revolutionized the study of cellular heterogeneity, enabling detailed molecular profiling at the individual cell level. However, integrating high-dimensional single-cell data into causal mediation analysis remains challenging due to zero inflation and complex mediator structures. We propose a novel mediation framework leveraging zero-inflated negative binomial models to characterize cell-level mediator distributions and beta regression for zero-inflation proportions. The model can identify expression level as well as expressed proportion that could mediate disease-leading causal pathway. Extensive simulation studies demonstrate improved power and controlled false discovery rates. We further illustrate the utility of this approach through application to ROSMAP single-cell transcriptomic data, uncovering biologically meaningful mediation effects that enhance understanding of disease mechanisms.}





\maketitle

\section{Introduction}\label{smore:introduction}
Single-cell RNA sequencing (scRNA-seq, hereafter) has transformed molecular biology by making it possible to measure transcriptomes at a scale and resolution that bulk assays cannot achieve. Single-cell assays provide gene expression profiles at the level of individual cells and uncover heterogeneity that is masked when averaging across cell populations. These data are high-dimensional and sparse, and two types of summaries are especially informative: the average expression across cells and the proportion of cells in which a gene is expressed (or equivalently the proportion of zeros). Most existing analyses \citep{startingref1, startingref2} have focused on average expression, but the expressed and non-expressed proportions capture different biological information related to prevalence and activation that have been largely overlooked. Current mediation methods for omics data do not account for proportion-based features or the sparsity inherent in single-cell measurements. To our knowledge, causal mediation analysis specifically tailored to single-cell data has not yet been explored in the published literature. 

Building on foundational work \citep{intro.seminal}, causal mediation analysis has become a widely applied framework for assessing the role and quantifying the effect of intermediate variables, known as mediators, that lie on the causal pathway from an exposure (e.g., disease status such as Alzheimer's disease versus healthy control) to an outcome (e.g., changes in laboratory measurements), within the counterfactual framework \citep{intro.counterfactual1}. This relationship is often conceptualized as exposure $\rightarrow$ mediator $\rightarrow$ outcome. Mediation analysis was originally more prevalent in the fields of social science and psychology \citep{intro.mediation1}, but it has since attracted growing attention from statisticians and has undergone substantial methodological advancements to address high-dimensional data with multiple mediators \citep{intro.mediation2, intro.mediation3, intro.mediation4}. This shift is especially relevant in modern biomedical research, where high-throughput technologies generate complex omics data that require mediation frameworks capable of adapting to the unique characteristics of each omics data type. For example, parametric linear structural equation modeling has been applied in sparse compositional mediation settings to estimate joint mediation effects \citep{microbiome1, quran}. Another line of work has focused on mediators selection using the isometric log-ratio transformation \citep{microbiome2}, as well as through linear log-contrast regression with Dirichlet regression in a two-part modeling framework to handle the high-dimensionality and zero-inflated structure commonly observed in microbiome data \citep{microbiome3}. Without compositionality as a factor to consider, epigenetic data such as DNA methylation are also high-dimensional. Several methods have been proposed to address key challenges, including bias correction in mediation estimation using the simulation extrapolation approach \citep{epigenetic1}, multiple testing procedures for high-dimensional mediators \citep{epigenetic2}, and mediator selection and effect estimation through a combination of sure independence screening, penalization, and the joint significance test \citep{epigenetic3, epigenetic4}. Similarly, transcriptomic data, typically generated through RNA-sequencing (RNA-Seq), are characterized by high-dimensionality and sparsity. A Cox model with random effects has been proposed to identify mediation effects of gene expression in the presence of high-dimensional exposures with a survival outcome \citep{transcriptomic1}. Although not focused on modeling the count distributions typical of RNA-Seq data, a recent study developed a meta-analysis framework that combines $R^{2}$-based total mediation effect estimates from multiple cohorts to identify mediator genes \citep{transcriptomic2}. However, these RNA-Seq data are based on bulk level measurements, capturing average gene expression across heterogeneous populations of cells \citep{bulkrnalimit}, which may obscure cell-specific effects and dilute signals of mediation.

Although there has been meaningful methodological progress across various omics platforms, including microbiome, DNA methylation, and transcriptomic data based on bulk RNA-Seq, a clear methodological gap remains. To our knowledge, no published work has addressed causal mediation analysis in the context of single-cell data, such as scRNA-seq, which provides the resolution needed to disentangle cell-type-specific mediation mechanisms. Accordingly, the goal of this paper is to develop a statistical methodology that accounts for the distinct characteristics of scRNA-seq data, with potential extension to single-nucleus RNA-Seq data. These data tend to show a high degree of cell-to-cell variability and a large number of zero gene counts, a pattern commonly referred to as zero inflation \citep{scrna1, zeroinf1, zeroinf2}. To address these challenges, we introduce a novel causal mediation framework based on beta and negative-binomial (NB) models that accommodates the challenging features of scRNA-seq data. 

We refer to this proposed method as \textbf{MedZIsc} (pronounced as ``Magics''), which stands for a statistical framework for co-\underline{Med}iators of \underline{Z}ero-\underline{I}nflated \underline{sc}RNA-Seq data. This name reflects its focus on modeling high-dimensional mediators while accounting for sparsity and cell-to-cell heterogeneity inherent to in single-cell transcriptomics data. The MedZIsc procedure consists of the following three main parts. First, we perform a preliminary screening of candidate mediator genes on the outcome variable using Lasso-penalized regression. For each gene, we separately fit marginal beta regression models and marginal NB regression models, thereby capturing the structural zero component and the nonzero count component of mediating pathway, respectively. Second, genes selected from the preliminary screening are incorporated into the outcome and mediator models to estimate interventional indirect effects (IIE) \citep{IIEref}. Specifically, we fit a linear regression model for the outcome, as well as marginal beta and marginal negative binomial regression models for each gene, following the model specifications described in Section \ref{smore:modelspec}. This allows us to estimate the IIE through expressed proportion and the IIE through average expression level. Third, we apply the joint significance (JS) test \citep{JS} to assess whether the IIEs for each gene are statistically significant, with appropriate control of the false discovery rate (FDR).

In this paper, we demonstrate the utility of MedZIsc through extensive simulation experiments and apply the method to scRNA-seq data from the Religious Orders Study and Rush Memory and Aging Project (ROSMAP) to uncover mediation pathways associated with neurodegenerative disease processes. The paper concludes with a summary of key findings, a discussion of limitations and challenges, and potential research avenues for future methodological development and biomedical applications.

\section{Methods}\label{smore:methods}
\subsection{Notations and Assumptions}
Let $Y$ be a continuous outcome, and let $X$ denote an exposure variable. Additionally, let $\bm{Z} = \{Z_{1}, \dots  Z_{K} \}$ represent covariates to be adjusted for, such as age, sex, BMI, and smoking history, from \textit{n} i.i.d. observations. For readability, subject-level subscripts are suppressed unless otherwise stated. Let $M_{cg}$ be the observed read count aligned to gene $g$ in cell $c$, for $g = 1, \dots , G$ and $c = 1, \dots , C$. scRNA-seq data often exhibit high variability and a substantial proportion of zero counts (i.e., zero-inflation) \citep{zeroinf1, zeroinf2}. To account for these characteristics, we incorporate two sets mediators within a counterfactual framework \citep{Rubin1, counterfactual1}. Specifically, let $M_{cg}(x)$ denote the potential read count of gene $g$ in cell $c$ under exposure level $X = x$. Let $I_{(M_{cg}(x) = 0)}$ indicate whether the gene has zero expression in cell $c$ (i.e., binary indicator of zero-inflation), capturing the proportion of cells with undetectable expression under exposure $X=x$.

The causal effects (i.e., IIEs) are assessed by taking the mean expected difference in counterfactual outcomes that would have been observed \citep{counterfactual2, counterfactual3, IIEref}. In this paper, we will assume that the IIEs are estimated under the assumptions of no-unmeasured confounders for the exposure-outcome, mediator-outcome, and exposure-mediator relationships. Notice that IIE estimation need less assumptions than estimating natural indirect effects. It does not require cross-world independence assumption between the mediator and outcome, and it does not need to assume any causal structures within the mediator to estimate IIE via a specific mediator \citep{IIEref}.

\subsection{Model Specifications}\label{smore:modelspec}

\begin{figure}[htb!]
\centering
\caption{A causal diagram to describe the framework of the mediation analysis involving two co-mediators, $M_{g}$ and $F_{g}$, between exposure $X$ and outcome $Y$. Here, two types of effects are assessed: (1) the interventional indirect effects between $X$ and $Y$, representing the mediation pathways through $M_{g}$ and $F_{g}$, respectively, and (2) the direct effect $\beta_{X}$ between exposure $X$ and outcome $Y$, which is not mediated by $M_{g}$ or $F_{g}$.}
\begin{tikzpicture}[node distance=2.5cm, thick, >=Stealth]
    \node (exposure) [draw, rectangle, rounded corners, align=center] {\textbf{Exposure} ($X$)};
    \node (outcome) [draw, rectangle, rounded corners, right=of exposure, align=center] {\textbf{Outcome} ($Y$)};
    \node (mediator) [draw, rectangle, rounded corners, above=1.8cm of $(exposure)!0.5!(outcome)$, align=center] {
        \textbf{Co-Mediators} \\
        $M_{1}, F_{1}$ \\
        $M_{2}, F_{2}$ \\
        $\vdots$ \\
        $M_{G}, F_{G}$
    }; 

    \draw[->] (exposure) -- (outcome) node[midway, below=2pt] {$\beta_{X}$}; 
    \draw[->] (exposure) -- (mediator) node[pos=0.5, above left=2pt] {$\alpha_{X}^{(g)}$ and $\gamma_{X}^{(g)}$}; 
    \draw[->] (mediator) -- (outcome) node[pos=0.5, above right=2pt] {$\beta_{M_{g}}$ and $\beta_{F_{g}}$}; 
\end{tikzpicture}
\label{smore:Fig1}
\end{figure}
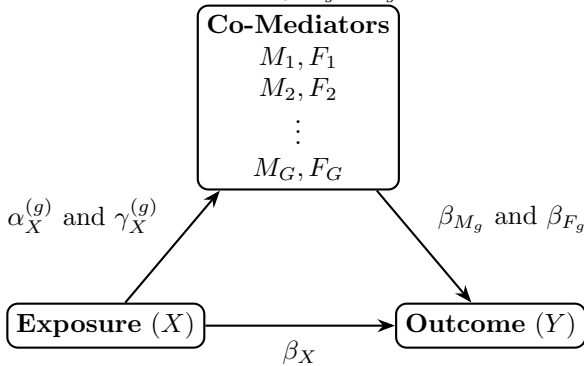

We propose the following models to describe the causal pathways described in Figure \ref{smore:Fig1}. The $M_{g}$ and $F_{g}$ are defined as $M_{g}= \frac{1}{C} \sum_{c=1}^{C} M_{cg}$ and $F_{g}= \frac{1}{C} \sum_{c=1}^{C} I(M_{cg} = 0)$ without loss of generality. The first model (Equation \ref{smore:outcomemodel}) is the outcome model, which describes the effect of mediator, decomposed into $M_{g}$ and $F_{g}$, on the relationship between exposure $X$ and outcome $Y$, while adjusting for covariates $\bm{Z}$. 

The second set of models includes two interconnected mediation models designed to account for the excess zeros and overdispersion commonly observed in scRNA-seq data. We model the gene expression count matrix, using a beta negative-binomial (BNB) framework. The BNB mediation model consists of two components: (1) a beta regression component modeling the probability of structural zeros, or $F_{g}$, which is the proportion of $M_{cg}$ that are zero across all $c$ (Equation \ref{smore:mediationmodel1}); and (2) a negative binomial (NB) regression component modeling the expected count given nonzero expression (Equation \ref{smore:mediationmodel2}). Notably, $F_{g}$ is a continuous response variable bounded in the interval $(0, 1)$, representing the non-expressed proportion, hence we assume it follows a beta distribution. 
\begin{align}
F_{g} \sim \text{Beta}(\mu_{F_{g}}, \phi)    \nonumber ,
\end{align}
\noindent where $\mu_{F_{g}}$ is the mean of $F_{g}$ and $\phi$ is a precision parameter of $F_{g}$. Additionally, the gene expression level $M_{g}$ are assumed to follow a NB distribution to account for overdispersion and excess zeros.
\begin{align}
M_{g} \sim \text{NB}(\mu_{M_{g}}, \theta)    \nonumber ,
\end{align}
\noindent where $\mu_{M_{g}}$ is the mean of $M_{g}$ and $\theta$ is the dispersion parameter of the NB distribution.
\begin{align}
 Y = \sum_{g=1}^{G}\beta_{M_{g}} M_{g}  & +  \sum_{g=1}^{G}\beta_{F_{g}} F_{g}  +  \beta_{X}X + \bm{\beta_{Z}}^\top \bm{Z} + \varepsilon, \label{smore:outcomemodel} \\ 
\text{logit}(\mu_{F_{g}}) &= \alpha_{X}^{(g)}X + \bm{\alpha_{Z}}^{(g)\top} \bm{Z}, \label{smore:mediationmodel1}\\
\log(\mu_{M_{g}}) &= \gamma_{X}^{(g)}X + \bm{\gamma_{Z}}^{(g)\top} \bm{Z} , \text{ for } g = 1, \dots, G,
\label{smore:mediationmodel2}
\end{align}

\noindent where $\beta_{M_{g}}$ represents the vector of coefficients for the mediator $g$th gene; $\beta_{F_{g}}$ denotes the vector of coefficients associated with the presence of nonzero gene expression; $\beta_{X}$ denotes the regression coefficient for the effect of $X$ on $Y$; $\bm{\beta_{Z}}^\top = (\beta_{Z_{1}}, \dots, \beta_{Z_{K}})$ is the vector of coefficients for the covariates; and $\epsilon$ is a normal random error variable in the outcome model. In the mediation models, $\alpha_{X}^{(g)}$ and $\gamma_{X}^{(g)}$ represent the effects of $X$ on the structural zero component and count component of gene $g$, respectively. $\bm{\alpha_{Z}}^{(g)\top} = (\alpha_{Z_{1}}^{(g)}, \dots, \alpha_{Z_{K}}^{(g)})$ and $\bm{\gamma_{Z}}^{(g)\top} = (\gamma_{Z_{1}}^{(g)}, \dots, \gamma_{Z_{K}}^{(g)})$ are the corresponding vectors of covariate effects. 

\subsection{Mediator Screening Procedure}\label{smore:screening}
Before fitting the final outcome and mediation models, we apply a screening procedure to identify candidate mediators. First, we fit the outcome model in Equation \ref{smore:outcomemodel} using Lasso regression, treating all aggregated mediators $M_{g}$ and $F_{g}$ as candidate predictors. This penalized approach selects a subset of genes whose expression level and expressed proportion are potentially associated with the outcome $Y$. Separately, we fit marginal models for each gene: a beta regression for $F_{g}$ (Equation \ref{smore:mediationmodel1}) and a NB regression for $M_{g}$ (Equation \ref{smore:mediationmodel2}). A gene is retained if it is selected by the Lasso model and exhibits a statistically significant association between the exposure $X$ and either $F_{g}$ or $M_{g}$ in its corresponding marginal model at 5$\%$ significance level. This criterion ensures that only genes supported by both the outcome model and at least one mediation pathway are included.

Let $\mathcal{G}_{Y}$, $\mathcal{G}_{M}$, and $\mathcal{G}_{F}$ denote the sets of gene indices selected from the outcome model, the $M_{g}$ model, and the $F_{g}$ model, respectively. The final set of candidate genes is given by the union  $\mathcal{S} = (\mathcal{G}_{Y} \cap \mathcal{G}_{M}) \cup (\mathcal{G}_{Y} \cap \mathcal{G}_{F}$). For each $g \in \mathcal{S}$, the term $M_{g}$ is included in the final outcome model if $g \in \mathcal{G}_{M}$, and $F_{g}$ is included if $g \in \mathcal{G}_{F}$. The complete set of predictors therefore consists of $X$, the covariates $\bm{Z}$, and the selected $M_{g}$ and $F_{g}$ terms.

\subsection{Hypothesis Testing}\label{smore:hypothesis}

From the model specification, the direct effect of $X$ on $Y$ (i.e., $X \rightarrow Y$) is estimated by the coefficient $\beta_{X}$ in the outcome model. We define two IIEs: (1) $\text{IIE}^{M_g}$, the indirect effect via $M_{g}$ (i.e., $X \rightarrow M_{g} \rightarrow Y$), and (2) $\text{IIE}^{F_g}$, the indirect effect via $F_{g}$ (i.e., $X \rightarrow F_{g} \rightarrow Y$). The IIE for $M_{g}$ when $X$ changes from $x_{1}$ to $x_{2}$ is:
\begin{align*}
    \text{IIE}^{M_g} = \E_{Z} \biggl[ Y\Bigl(x_{2}, W_{M_{g}}(x_{2}), W_{-M_{g}}(x_{1})\Bigr) - Y\Bigl(x_{2}, W_{M_{g}}(x_{1}), W_{-M_{g}}(x_{1})\Bigr) \biggr],
\end{align*}
where $W_{M_{g}}(x_{1})$ and $W_{M_{g}}(x_{2})$ denote a random draw from the distribution of $M_{g}(x_{1})$ and $M_{g}(x_{2})$, respectively, and $W_{-M_{g}}(x_{1})$ denotes a random draw from the multivariate distribution of $\Bigl( M_{1}(x_{1}),\allowbreak F_{1}(x_{1}),\allowbreak M_{2}(x_{1}),\allowbreak F_{2}(x_{1}),\allowbreak \cdots,\allowbreak M_{g-1}(x_{1}),\allowbreak F_{g-1}(x_{1}),\allowbreak F_{g}(x_{1}),\allowbreak M_{g+1}(x_{1}),\allowbreak F_{g+1}(x_{1}),\allowbreak \cdots,\allowbreak M_{G}(x_{1}),\allowbreak F_{G}(x_{1}) \Bigr)$. Thus, we have
\begin{align*}
    \text{IIE}^{M_g} &= \beta_{X}x_{2} + \beta_{Z}^{\top}Z + \beta_{M_g}\E_{Z} \Bigl(W_{M_{g}}(x_{2})\Bigr) + \sum_{i \neq g}^{G} \beta_{M_i}\E_{Z} \Bigl(W_{M_{i}}(x_{1})\Bigr)  \\
    & \qquad\qquad + \sum_{g = 1}^{G} \beta_{F_g}\E_{Z} \Bigl(W_{F_{g}}(x_{1})\Bigr) - \beta_{X}x_{2} - \beta_{Z}^{\top}Z - \beta_{M_g}\E_{Z} \Bigl(W_{M_{g}}(x_{1})\Bigr) \\
    & \qquad\qquad - \sum_{i \neq g}^{G} \beta_{M_i}\E_{Z} \Bigl(W_{M_{i}}(x_{1})\Bigr) - \sum_{g = 1}^{G} \beta_{F_g}\E_{Z} \Bigl(W_{F_{g}}(x_{1})\Bigr) \\
    &= \beta_{M_g}\E_{Z} \Bigl(W_{M_{g}}(x_{2})\Bigr) - \beta_{M_g}\E_{Z} \Bigl(W_{M_{g}}(x_{1})\Bigr) \\
    &= \beta_{M_g} \Bigl( \exp(\gamma_{X}^{(g)}x_{2} + \gamma_{Z}^{(g)}Z) - \exp(\gamma_{X}^{(g)}x_{1} + \gamma_{Z}^{(g)}Z)  \Bigr) \\
    &= \beta_{M_g} \exp(\gamma_{Z}^{(g)}Z) \Bigl( \exp(\gamma_{X}^{(g)}x_{2}) - \exp(\gamma_{X}^{(g)}x_{1})  \Bigr).
\end{align*}
Similarly, the IIE for $F_{g}$ when $x$ changes from $x_{1}$ to $x_{2}$ is:
\begin{align*}
    \text{IIE}^{F_g} = \E_{Z} \biggl[ Y\Bigl(x_{2}, W_{F_{g}}(x_{2}), W_{-F_{g}}(x_{1})\Bigr) - Y\Bigl(x_{2}, W_{F_{g}}(x_{1}), W_{-F_{g}}(x_{1})\Bigr) \biggr],
\end{align*}
where $W_{F_{g}}(x_{1})$ and $W_{F_{g}}(x_{2})$ denote a random draw from the distribution of $F_{g}(x_{1})$ and $F_{g}(x_{2})$, respectively, and $W_{-F_{g}}(x_{1})$ denotes a random draw from the distribution of the vector $\Bigl( M_{1}(x_{1}),\allowbreak F_{1}(x_{1}),\allowbreak M_{2}(x_{1}),\allowbreak F_{2}(x_{1}),\allowbreak \cdots,\allowbreak M_{g}(x_{1}),\allowbreak M_{g+1}(x_{1}),\allowbreak F_{g+1}(x_{1}),\allowbreak \cdots,\allowbreak M_{G}(x_{1}),\allowbreak F_{G}(x_{1}) \Bigr)$. Thus, we have
\begin{align*}
    \text{IIE}^{F_g} &= \beta_{X}x_{2} + \beta_{Z}^{\top}Z +  \sum_{g = 1}^{G} \beta_{M_g}\E_{Z} \Bigl(W_{M_{g}}(x_{1})\Bigr) + \beta_{F_g}\E_{Z} \Bigl(W_{F_{g}}(x_{2})\Bigr) \\
    & \qquad\qquad + \sum_{i \neq g}^{G} \beta_{F_g}\E_{Z} \Bigl(W_{F_g}(x_{1})\Bigr) - \beta_{X}x_{2} - \beta_{Z}^{\top}Z - \sum_{g = 1}^{G} \beta_{M_g}\E_{Z} \Bigl(W_{M_{g}}(x_{1})\Bigr) \\
    & \qquad\qquad - \beta_{F_g}\E_{Z} \Bigl(W_{F_{g}}(x_{1})\Bigr) - \sum_{i \neq g}^{G} \beta_{F_g}\E_{Z} \Bigl(W_{F_{g}}(x_{1})\Bigr)  \\
    &= \beta_{F_g} \Bigl( \E_{Z} \Bigl(W_{F_{g}}(x_{2})\Bigr) - \E_{Z} \Bigl(W_{F_{g}}(x_{1})\Bigr) \Bigr) \\
    &= \beta_{F_g} \Bigl( \expit(\alpha_{X}^{(g)}x_{2} + \alpha_{Z}^{(g)}Z) - \expit(\alpha_{X}^{(g)}x_{1} + \alpha_{Z}^{(g)}Z)  \Bigr),
\end{align*}
where $\expit(\cdot)$ is the inverse of $\logit(\cdot)$ function.

To formally assess these mediation effects, we construct the following hypothesis testing framework for each gene $g$:
\begin{align}
H_{0}^{M}: \beta_{M_{g}}\gamma_{X}^{(g)} &= 0, \nonumber \\
H_{0}^{F}: \beta_{F_{g}}\alpha_{X}^{(g)} &=0 \nonumber.
\end{align}
Under $H_{0}^{M}$, either $\beta_{M_{g}}$ or $\gamma_{X}^{(g)}$ is zero, implying no mediation through $M_{g}$. Similarly, under $H_{0}^{F}$, either $\beta_{F_{g}}$ or $\alpha_{X}^{(g)}$ is zero, implying no mediation through $F_{g}$. The joint significance (JS) test \citep{JS} is used to assess whether a gene mediates the causal effect of $X$ on $Y$ through its expression level or the proportion of expressing cells. The JS test is known to control the type I error rate while maintaining statistical power in various omics studies \citep{JS.study1, JS.study2}. To test $H_{0}^{M}$ and $H_{0}^{F}$, we use the maximum of two p-values from the relevant coefficient estimates. Specifically, we define
\begin{align*}
    P_{\max_{g}}^{M} &= \max(P_{\beta_{M_{g}}}, P_{\gamma_{X}^{(g)}}), \nonumber \\
    P_{\max_{g}}^{F} &= \max(P_{\beta_{F_{g}}}, P_{\alpha_{X}^{(g)}}),
\end{align*}
where $P_{\beta_{M_{g}}}$ and $P_{\beta_{F_{g}}}$ are p-values obtained from the outcome model (Equations \ref{smore:outcomemodel}), and $P_{\alpha_{X}^{(g)}}$ and $P_{\gamma_{X}^{(g)}}$ are p-values obtained from the mediator models (Equations \ref{smore:mediationmodel1} and \ref{smore:mediationmodel2}, respectively). Of note, $P_{\max_{g}}^{M}$ and $P_{\max_{g}}^{F}$ are used to construct the heatmap. Each p-value is calculated using the following test statistic:
\begin{align*}
    P_{\beta_{M_{g}}} &= 2\biggl\{1 - F\biggl(\frac{|\hat{\beta}_{M_{g}}|}{\hat{\sigma}_{\beta_{M_{g}}}}\biggr) \biggr\}, \nonumber \\
    P_{\beta_{F_{g}}} &= 2\biggl\{1 - F\biggl(\frac{|\hat{\beta}_{F_{g}}|}{\hat{\sigma}_{\beta_{F_{g}}}}\biggr) \biggr\}, \nonumber \\
    P_{\gamma_{X}^{(g)}} &= 2\biggl\{1 - F\biggl(\frac{|\hat{\beta}_{\gamma_{X}^{(g)}|}}{\hat{\sigma}_{\gamma_{X}^{(g)}}}\biggr) \biggr\}, \nonumber \\
    P_{\alpha_{X}^{(g)}} &= 2\biggl\{1 - F\biggl(\frac{|\hat{\beta}_{\alpha_{X}^{(g)}|}}{\hat{\sigma}_{\alpha_{X}^{(g)}}}\biggr) \biggr\}, \nonumber
\end{align*}
where $\hat{\beta}_{M_{g}}$, $\hat{\beta}_{F_{g}}$, $\hat{\beta}_{\gamma_{X}^{(g)}}$, and $\hat{\beta}_{\alpha_{X}^{(g)}}$ are the estimated regression coefficients from the fitted outcome and mediation models based on Equations \ref{smore:outcomemodel} to \ref{smore:mediationmodel2}, and $\hat{\sigma}_{\beta_{M_{g}}}$, $\hat{\sigma}_{\beta_{F_{g}}}$, $\hat{\sigma}_{\gamma_{X}^{(g)}}$, and $\hat{\sigma}_{\alpha_{X}^{(g)}}$ are their corresponding standard error estimates. The function $F(\cdot)$ denotes the cumulative distribution function of the standard normal distribution. Given the large number of hypothesis tests across genes, it is important to control the false discovery rate (FDR). In this study, we apply the BH-adjusted p-value \citep{BH} to the p-values of the selected mediators, using \texttt{stats R} package.

\subsection{Handling Special Cases in Mediation Model}\label{smore:zerohandling}
In our proposed mediation model, $F_g$ represents the proportion of zero counts for gene $g$ across all cells. Since $F_g$ is a proportion, it will fall between 0 and 1. To handle edge cases where $F_g$ is exactly 0 or 1, we make the following adjustments below.

\begin{enumerate}
    \item[\noindent(i)] \textbf{$F_{g}$ is 0 for some subjects:} This happens when a gene is expressed in all cells for some subjects (i.e., $M_{cg} > 0$ for all $c$), resulting in $F_{g} = 0$ for those subjects. Having $F_{g}$ exactly equal to 0 can lead to computational issues in the beta regression modeling of the BNB mediation model (Equation \ref{smore:mediationmodel1}) due to the bounded nature of beta distribution. To address this, we apply the following adjustment:
    \begin{align*}
        F_{g} = \max(F_{g}, 0.001).
    \end{align*}
    \item[\noindent(ii)] \textbf{$F_{g}$ is 0 across all subjects:} This happens when a gene is expressed in all cells of all subjects. If $F_g$ is exactly 0 across all subjects for a given gene $g$, the variability required to model $F_{g}$ is lost. In this case, we modify the mediation modeling procedure by omitting Equation \ref{smore:mediationmodel1} for the affected gene and removing the $F_{g}$ terms from the outcome model (Equation \ref{smore:outcomemodel}).
    \item[\noindent(iii)] \textbf{$F_{g}$ is 1 for some subjects:} This occurs when a gene is not expressed in all cells for some subjects (i.e., $M_{cg} = 0$ for all $c$), resulting in $F_{g} = 1$. As the upper bound of the beta distribution is 1, values near this boundary may introduce numerical issues. Subsequently, we make the following adjustment to prevent this special case:
    \begin{align*}
        F_{g} = \min(F_{g}, 0.999).
    \end{align*}
\end{enumerate}

\subsection{Performances Measures}
In this study, we evaluate the performance of the proposed method using statistical power and FDR, averaging across simulation replicates. For each simulation run, the interventional-specific JS test is applied to each gene, separately assessing $M_{g}$ and $F_{g}$. Mediators are identified at a $5\%$ significance level after the BH multiplicity correction. A gene is considered a true positive if it is found among simulated mediators and is declared statistically significant. Statistical power is defined as the proportion of true mediators that are correctly identified: 
\begin{align*}
    \text{Power} &= \frac{\text{Number of True Positives}}{\text{Number of True Mediators}}. \nonumber 
\end{align*}
FDR is calculated as the proportion of false positives among all genes declared significant (or discoveries):
\begin{align*}
    \text{FDR} &= \frac{\text{Number of False Positives}}{\text{Number of Discoveries}} \cdot \mathbbm{1}{\{\text{Number of Discoveries} > 0\}}. \nonumber 
\end{align*}

\section{Simulation Study}\label{smore:results}
\subsection{Simulation Design}\label{smore:sim.design}
In this study, we generated data to evaluate the performance of the proposed method within a causal mediation framework for scRNA-seq data. Scenarios I-III varied in sample size $n = 100, 400, 800$, number of cells per subject $c=100, 200$, and number of genes $g=100, 200, 400$. For Scenario IV, we adopted a real-data driven configuration to reflect large-scale experiments, setting $n=400$ subjects, $c=2000, 3000$ cells per subject, and $g=20000, 30000$ genes. The binary exposure variable $X$ was drawn from a Bernoulli distribution with probability of 0.5, and covariates $\bm{Z} \in \mathbb{R}^3$ were generated from a standard normal distribution. 

For each subject $i$ and gene $g$, cell-level mediator gene expression counts $M_{cg}^{(i)}$ were simulated from a zero-inflated negative binomial (ZINB) distribution with gene-specific mean $\mu_{ig}$, zero-inflation probability $\pi_{ig}$, and dispersion $\delta_{g}$, to better mimic the excess zeros nature of scRNA-seq data. These parameters were defined as follows:
\begin{align}
    \text{logit}(\pi_{ig}) &= \alpha_{X}^{(g)}X_{i} + \bm{\alpha_{Z}}^{(g)\top} \bm{Z}_{i}, \label{smore:meddatagen1} \\
    \log(\mu_{ig}) &= \gamma_{X}^{(g)}X_{i} + \bm{\gamma_{Z}}^{(g)\top} \bm{Z}_{i}, \label{smore:meddatagen2}
\end{align}
where $\mu_{ig}$ and $\pi_{ig}$ are subject-level parameters capturing the expected gene expression and zero-inflation probability, respectively.

Using these equations, we first determine whether a gene expression was a structural zero, based on the binomial random variable with probability $\pi_{ig}$ (see Equation \ref{smore:meddatagen1}). If not, expressions were drawn from a NB distribution with mean $\mu_{ig}$ (see Equation \ref{smore:meddatagen2}) and each gene was assigned a random dispersion parameter $\delta_{g} \sim \mathrm{Unif}(0.6, 1.2)$. 

This yields the ZINB cell-level expression, which we define as:
\begin{align*}
    M_{cg}^{(i)} \sim ZINB(\mu_{ig}, \delta_{g}, \pi_{ig}).
\end{align*}
We then aggregate the cell-level measurements for each subject by taking the average of the cell-level $M_{cg}$:
\begin{align*}
    M_{g}^{(i)} &= \frac{1}{C} \sum_{c = 1}^{C} M_{cg}^{(i)}, \\
    F_{g}^{(i)} &= \frac{1}{C} \sum_{c = 1}^{C} I(M_{cg}^{(i)} = 0),
\end{align*}
where $M_{g}^{(i)}$ represents the average expression and $F_{g}^{(i)}$ represents the proportion of zeros (i.e., non-expressed proportion) for gene $g$ in subject $i$.

To ensure biologically plausability in the simulated data, we incorporated additional data pre-processing as described in Section \ref{smore:zerohandling}. Specifically, genes with no expression across all subjects were removed from the analysis, proportions of zeros were bounded away from 0 and 1 by:
\begin{align*}
    \hat{F}_{g}^{(i)} = \min(\max(F_{g}^{(i)}, 0.001), 0.999).    
\end{align*}
This was necessary to be informative for the downstream modeling. Among all genes, we randomly selected a subset of true mediators, which were divided into three groups: (1) both-path mediators affected by exposure and contributing to the outcome through both $M_{g}$ and $F_{g}$, (2) $M$-only mediators contributing through $M_{g}$ only, (3) $F$-only mediators contributing through $F_{g}$ only. Approximately half were designatured as both-path mediators, with the remaining split between $M$-only and $F$-only groups. For each true mediator gene $g \in \mathcal{G}_{\text{true}}$, the coefficients for the exposure effects on the mediator components were generated as follows: for both-path mediators, $\alpha_{X}^{(g)} \sim \mathrm{Unif}(1, 2)$ and $\gamma_{X}^{(g)} \sim \mathrm{Unif}(2, 6)$; for $M$-only mediators, $\alpha_{X}^{(g)} \sim \mathrm{Unif}(1, 1.5)$; and for $F$-only mediators, $\gamma_{X}^{(g)} \sim \mathrm{Unif}(1.8, 3)$. Covariate effects on the mediators were fixed at $\alpha_{Z} = 0.1$ and $\gamma_{Z} = 0.3$ across all genes and covariates, corresponding to Equations \ref{smore:meddatagen1} and \ref{smore:meddatagen2}.

The outcome variable $Y$ was simulated from a linear regression model including $X$, $\bm{Z}$, and aggregated mediator features  $M_{g}$ and $F_{g}$ for all $g$. The coefficients in the outcome model were set to be $\beta_{X} = 3$ and $\bm{\beta_{Z}} = (0.5, -0.3, 0.2)$. For both-path mediators, $\beta_{M_{g}} \sim \mathrm{Unif}(4, 5)$ and $\beta_{F_{g}} \sim \mathrm{Unif}(12, 14)$; for $M$-only mediators, $\beta_{M_{g}} \sim \mathrm{Unif}(5, 8)$; and for $F$-only mediators, $\beta_{F_{g}} \sim \mathrm{Unif}(10, 15)$. A normally distributed error term was added to the outcome. 

For benchmarking the performance of MedZIsc, a naïve approach (i.e., marginal modeling) was also included. In this setting, each separate gene was regressed separately on the outcome along with $X$ and $\bm{Z}$, representing for the marginal outcome model, and each gene is also regressed as a function of the $X$ and $\bm{Z}$ for the marginal mediation model. No preliminary screening is applied in the naïve approach, but the same JS-test with BH-adjustment used. All simulation results are averaged over 100 replicates. 

\subsection{Simulation Results}\label{smore:sim.results}
In our simulation experiments, we evaluated performance across four scenarios spanning simple to large-scale designs. Table \ref{smore:table1} summarizes scenarios I–III, which involved small to moderate to large sample sizes ($n = 100, 400, 800$), varying numbers of cells per subject ($c = 100, 200$), and gene sizes ($g = 100, 200, 400$). Scenario IV was designed to mimic real-data conditions, and is further detailed in Table \ref{smore:table2}.

As shown in Table 1, MedZIsc consistently achieved high power in assessing $\text{IIE}^{M_g}$ (denoted as Power(M) in the table) through the NB component (M model; see Equation \ref{smore:mediationmodel2}) of the BNB framework, while adequately controlling the FDR at the nominal 5$\%$ level. In Scenario I, the naïve approach had lower power than MedZIsc when the sample size was relatively small ($n=100$), but surpassed MedZIsc in power as the sample size increased in Scenarios II and III. However, this gain in power for the naïve approach comes at a cost, as its inflated FDR suggests a high rate of false discoveries in identifying $\text{IIE}^{M_g}$.

Furthermore, across all three scenarios in Table 1, MedZIsc demonstrated strong performance in detecting $\text{IIE}^{F_g}$ (denoted as Power(F) in the table) via the beta regression component (F model; see Equation \ref{smore:mediationmodel1}) of the BNB framework, while maintaining FDR control near the nominal 5$\%$ level. In Scenario I, MedZIsc generally yielded higher power than the naïve approach, except in a few configurations where $c = 200$ and either $c = g$ or $c < g$. The naïve approach, on the other hand, frequently showed substantial FDR inflation, though it remained below the threshold in limited cases such as $c = g = 100$ and $c < g = 200$ when $c = 100$. As sample size increased in Scenarios II and III, MedZIsc continued to exhibit favorable results, with both high detection rates and stable FDR. Although the naïve approach occasionally achieved similar power, it consistently failed to control the FDR.

Table \ref{smore:table2} summarizes results for Simulation IV, which was constructed to emulate the complexity of real-data application. This setting incorporated a larger sample size and substantially increased dimensionality ($n = 400$, $c = 2000, 3000$, and $g = 20000, 30000$). As a whole, the naïve approach yielded strong performance in terms of power across both $M$ and $F$ models. MedZIsc also retained high power for detecting $M$ mediators, though power for $F$ mediators declined under these realistic conditions. Nonetheless, MedZIsc consistently maintained FDR control near the nominal 5$\%$ level for both components. In contrast, the naïve method's high power was again accompanied by inflated FDR, which mirrors trends seen in Scenarios I–III (Table \ref{smore:table1}) and raises concerns about its reliability under high-dimensional configurations.

MedZIsc generally completed the simulation experiments faster than the naïve method, as summarized by the average computation time in both tables. This computational efficiency is expected, as MedZIsc applies penalized and marginal models during screening, but limits hypothesis testing to a subset of genes that are deemed relevant. All simulations were run on the Minerva supercomputing cluster at the Icahn School of Medicine at Mount Sinai using 5 CPU cores and 8 GB of RAM per node.

\begin{table*}[!htb]
\centering
\small
\setlength{\tabcolsep}{3pt}
\caption{Simulation results comparing the MedZIsc approach with the naïve approach across Scenarios I–III.}
\label{smore:table1}
\begin{tabular}{@{} ccc cccccc @{}}
\toprule
 $n$ & $c$ & $g$ & Methods & Power(M) & Power(F) & FDR(M) & FDR(F) & \makecell{Average \\ Comp. \\ Time \\ (in secs)} \\
\midrule 
\multicolumn{9}{@{}l}{Scenario I} \\
 100 & 100 & 100 & MedZIsc & 0.990 & 0.880 & 0.023 & 0.018 & 1.28 \\
  &  &  & Naïve & 0.585 & 0.352 & 0.462 & 0.038 & 1.64 \\
   & & 200 & MedZIsc & 0.993 & 0.836 & 0.046 & 0.023 & 2.47  \\
     &  &  & Naïve & 0.740 & 0.520 & 0.399 & 0.033 & 3.33 \\
   && 400 & MedZIsc &  0.957 & 0.824 & 0.031 & 0.035 & 4.91 \\
     &  &  & Naïve & 0.658 & 0.603 & 0.431 & 0.096 & 6.71 \\
  & 200 & 100 & MedZIsc & 0.990 & 0.812 & 0.021 & 0.008 & 1.38 \\
    &  &  & Naïve & 0.720 & 0.776 & 0.409 & 0.135 & 1.91 \\
   & & 200 & MedZIsc & 0.993 & 0.720 & 0.029 & 0.012 & 2.72  \\
     &  &  & Naïve & 0.750 & 0.868 & 0.400 & 0.141 & 3.89 \\
    && 400 & MedZIsc & 0.923 & 0.673 & 0.049 & 0.038 & 5.51  \\
      &  &  & Naïve & 0.665 & 0.991 & 0.429 & 0.191 & 8.00 \\
\midrule
\multicolumn{9}{@{}l}{Scenario II} \\
 400 & 100 & 100 & MedZIsc & 0.992 & 0.890 & 0.023 & 0.013 & 3.13 \\
   &  &  & Naïve & 1.000 & 0.869 & 0.250 & 0.280 & 3.91 \\
  & & 200 & MedZIsc & 0.993 & 0.868 & 0.028 & 0.017 & 5.86  \\
    &  &  & Naïve & 0.997 & 0.825 & 0.251 & 0.286 & 7.63 \\
     & & 400 & MedZIsc & 0.997 & 0.937 & 0.019 & 0.020 & 11.5 \\
       &  &  & Naïve & 1.000 & 0.911 & 0.250 & 0.266 & 15.4 \\
 & 200 & 100 & MedZIsc & 0.968 & 0.919 & 0.009 & 0.019 & 3.66 \\
   &  &  & Naïve & 1.000 & 0.891 & 0.250 & 0.274 & 4.70 \\
      & & 200 & MedZIsc & 0.973 & 0.895 & 0.013 & 0.014 & 6.93  \\
        &  &  & Naïve & 1.000 & 0.872 & 0.250 & 0.278 & 9.85 \\
      && 400 & MedZIsc & 0.981 & 0.953 & 0.016 & 0.015 & 13.6  \\
        &  &  & Naïve & 1.000 & 0.934 & 0.250 & 0.267 & 18.7 \\
\midrule
\multicolumn{9}{@{}l}{Scenario III} \\
 800 & 100 & 100 & MedZIsc & 0.960 & 0.930 & 0.017 & 0.008 & 5.57 \\
   &  &  & Naïve & 1.000 & 0.913 & 0.250 & 0.272 & 6.53 \\
     & & 200 & MedZIsc & 0.948 & 0.922 & 0.033 & 0.020 & 10.8  \\
       &  &  & Naïve & 1.000 & 0.890 & 0.250 & 0.276 & 13.3 \\
      & & 400 & MedZIsc & 0.964 & 0.959 & 0.020 & 0.013 & 21.3  \\
        &  &  & Naïve & 1.000 & 0.938 & 0.250 & 0.263 & 26.0 \\
   & 200 & 100 & MedZIsc & 0.894 & 0.943 & 0.014 & 0.015 & 7.23  \\
     &  &  & Naïve & 1.000 & 0.933 & 0.250 & 0.271 & 10.3 \\
      & & 200 & MedZIsc & 0.893 & 0.932 & 0.015 & 0.014 & 14.0  \\
        &  &  & Naïve & 1.000 & 0.911 & 0.250 & 0.275 & 16.6 \\
      && 400 & MedZIsc & 0.928 & 0.978 & 0.021 & 0.014 & 27.1  \\
        &  &  & Naïve & 1.000 & 0.965 & 0.250 & 0.264 & 33.6 \\
\bottomrule
\end{tabular}
\begin{tablenotes}
\item [] Abbreviations: Power(M), power for M models; Power(F), power for F models; FDR(M), false discovery rate for M models; FDR(F), false discovery rate for F models.
\item [] $n$ = sample size; $c$ = cells per subject; $g$ = number of genes.
\end{tablenotes}
\end{table*}

\begin{table*}[!htb]
\centering
\small
\setlength{\tabcolsep}{3pt}
\caption{Simulation results comparing the MedZIsc approach with the naïve approach under Scenario IV, reflecting real-data-like parameter configurations.}
\label{smore:table2}
\begin{tabular}{@{} ccc cccccc @{}}
\toprule
 $n$ & $c$ & $g$ & Methods & Power(M) & Power(F) & FDR(M) & FDR(F) & \makecell{Average \\ Comp. \\ Time \\ (in secs)} \\
\midrule
\multicolumn{9}{@{}l}{Scenario IV} \\
 400 & 2000 & 20000 & MedZIsc & 0.870 & 0.524 & 0.049 & 0.068 & 4116.9 \\
 &  &  & Naïve & 1 & 0.863 & 0.250 & 0.286 & 6136.7 \\
      & & 30000 & MedZIsc & 0.900 & 0.525 & 0.049 & 0.068 & 10711.2 \\
      &  &  & Naïve & 0.898 & 0.947 & 0.271 & 0.265 & 8112.9 \\
   & 3000 & 20000 & MedZIsc & 0.823 & 0.324 & 0.029 & 0.052 & 5901.3 \\
   &  &  & Naïve & 1 & 0.861 & 0.250 & 0.281 & 5480.5 \\
      & & 30000 & MedZIsc & 0.834 & 0.304 & 0.037 & 0.054 & 10853.9 \\
      &  &  & Naïve & 0.895 & 0.950 & 0.272 & 0.266 & 9806.8 \\
\bottomrule
\end{tabular}
\begin{tablenotes}
\item [] Abbreviations: Power(M), power for M models; Power(F), power for F models; FDR(M), false discovery rate for M models; FDR(F), false discovery rate for F models.
\item [] $n$ = sample size; $c$ = cells per subject; $g$ = number of genes.
\end{tablenotes}
\end{table*}

\section{Real Data Application}
\subsection{ROSMAP Study and Data Description}
Alzheimer's disease (AD) is a progressive neurodegenerative disorder and a leading cause of dementia in older adults. While its clinical and pathological features are well described, the mechanisms underlying cognitive decline or preserved function in the presence of AD pathology remain unclear. Recent single-cell transcriptomic studies from the Religious Orders Study and the Rush Memory and Aging Project (ROSMAP) cohort \citep{ROSMAPoriginal} have begun to address these questions by profiling over two million nuclei from postmortem prefrontal cortex samples of 427 study participants. These studies examined how different cell types and molecular features relate to cognitive impairment, AD pathology, and cognitive resilience. They identified selectively vulnerable neuronal subtypes and reported changes in glial, vascular, and epithelial cell composition, along with gene expression programs related to DNA damage and neuronal survival.

In particular, we are motivated by the potential role of vascular and epithelial cells. A recent single-cell study using AD-affected brain regions from the ROSMAP cohort reported increased abundance of these cells in late-stage AD, especially in the entorhinal cortex, hippocampus, and prefrontal cortex \citep{ROSMAP1}. Vascular and epithelial cells account for only about $0.8\%$ of all cells in another study that profiled seven major cell types from the same cohort \citep{ROSMAPdatapaper}. Although they are relatively rare, cerebrovascular cells are important for maintaining brain homeostasis. Some studies have pointed out that these cell types are still understudied, partly because of their sparsity and anatomical dispersion \citep{ROSMAP2}. There is also growing evidence that cerebrovascular dysfunction may be associated with AD pathology and early cognitive decline \citep{ROSMAP3}.

In our analysis, postmortem interval (PMI) is treated as the outcome variable. PMI is defined as the time between the recorded time of death and the time at which brain tissue is collected and stabilized through freezing or fixation. Although PMI is not a clinical endpoint in the traditional sense, in our framework it is treated as a biologically relevant outcome that may serve as a proxy for processes unfolding during the agonal and immediate postmortem period. Variability in PMI may reflect underlying differences in physiological stress, inflammation, agonal state, or pre-terminal brain pathology \citep{postmortem1, postmortem2}. For example, individuals experiencing greater systemic stress or disrupted cerebral homeostasis at the time of death may exhibit molecular or cellular changes that influence tissue stability and degradation, ultimately affecting the length of the postmortem interval. In this sense, PMI may capture meaningful biological variation tied to the condition of the body and brain around the time of death, rather than functioning solely as a technical variable or a marker of tissue quality. We are interested in whether AD status is associated with PMI, and whether this association is mediated by vascular or epithelial cell composition, particularly in terms of aggregated abundance and sparsity patterns. Sex was included as an additional covariate. In the ROSMAP dataset, both the pathologic diagnosis of Alzheimer's disease and the single-cell RNA-seq measurements were obtained at or shortly after the time of death, which is also the starting point for calculating the postmortem interval (PMI) \citep{ROSMAPdatapaper, ROSMAPtiming}. Since these data are derived from postmortem tissue, the timing of the exposure and mediator is aligned with the beginning of PMI, making it reasonable to treat PMI as the outcome. The analysis uses pre-processed single-cell gene count data for vascular and epithelial cells (provided as \texttt{Vasculature\_cells.rds}) and clinical metadata (provided as \texttt{individual\_metadata\_deidentified.tsv}) from 427 participants from the ROSMAP study. These data were obtained from the AD/Aging Brain Atlas (\url{https://compbio.mit.edu/ad_aging_brain/}) and had been log-normalized by the original authors using Seurat's \texttt{LogNormalize} method. The gene count matrix (33,538 genes by 17,974 cells) was loaded and processed as a Seurat object using the \texttt{Seurat R} package. It included 423 participants, whereas the metadata listed 427 participants. Four participants without corresponding gene count data were removed from the metadata. One additional participant was excluded due to missing PMI, resulting in a final sample size of 422 participants ($n = 237$ and $n = 186$ with and without a pathologic diagnosis of AD, respectively). Before applying MedZIsc, we performed gene filtering by removing genes with zero expression across all subjects, retaining only those expressed in at least one individual. This step reduced the number of genes from 33,538 to 27,516 for analysis. To meet the requirement of beta regression that the outcome lies strictly within the open interval $(0, 1)$, we adjusted the zero proportion values $F_{g}$ by bounding values of 0 and 1 to 0.001 and 0.999, respectively. Additionally, for each subject, we computed the average gene expression ($M_{g}$) and the proportion of zero counts ($F_{g}$) across cells, consistent with our methodology and simulation design. Although the reference dataset paper \citep{ROSMAPdatapaper} did not compute $M_{g}$ and $F_{g}$ in exactly the same way, it similarly summarized cell-level expression similar to our approach.

\subsection{Analysis of the ROSMAP Dataset}
Upon applying the proposed MedZIsc procedure to the ROSMAP dataset, we identified 1 out of 27,516 genes as significant $M$-mediators based on the $\text{IIE}^{M_g}$ estimates from the $M$ model, and 14 genes as significant $F$-mediators based on the $\text{IIE}^{F_g}$ estimates from the $F$ model. The complete lists of selected mediator genes from each model are provided in Table \ref{smore:rosmaptable}. One gene, NFKBIA, was identified by both the $M$ and $F$ models when using a $10\%$ significance level. The estimated pathway effect along $X \rightarrow M$ and $M \rightarrow Y$ are 4.466 and 0.234, resulting in a positive IIE via M model ($p_{max}$ = 0.055). In contrast, a negative IIE ($p_{max}$ = 0.018) is observed via the F model, based on the pathway effects of 12.80 and -0.283 between $X \rightarrow F$ and $F \rightarrow Y$, respectively. An important finding is that not only expression levels captured by the M model, but also the non-expressed proportions from the F model, can serve as mediators along the causal pathway. The estimated IIEs showed opposite directions in the two models, positive for M and negative for F. This is a novel feature of our approach, suggesting that the proportion of zeros may provide additional insight in single-cell analyses beyond what is explained by average expression. NFKBIA is a core regulatory component of the Nuclear Factor Kappa-b pathway, which has been implicated in the pathogenesis of Alzheimer's disease through its roles in neuronal survival, synaptic plasticity, memory regulation, microglial activity, and inflammation \citep{NFKBIA1}. Variants in NFKBIA have also been reported as potential risk factors for late-onset AD \citep{NFKBIA2}.

\begin{figure}[!ht] 
\centering
\includegraphics[scale=.44]{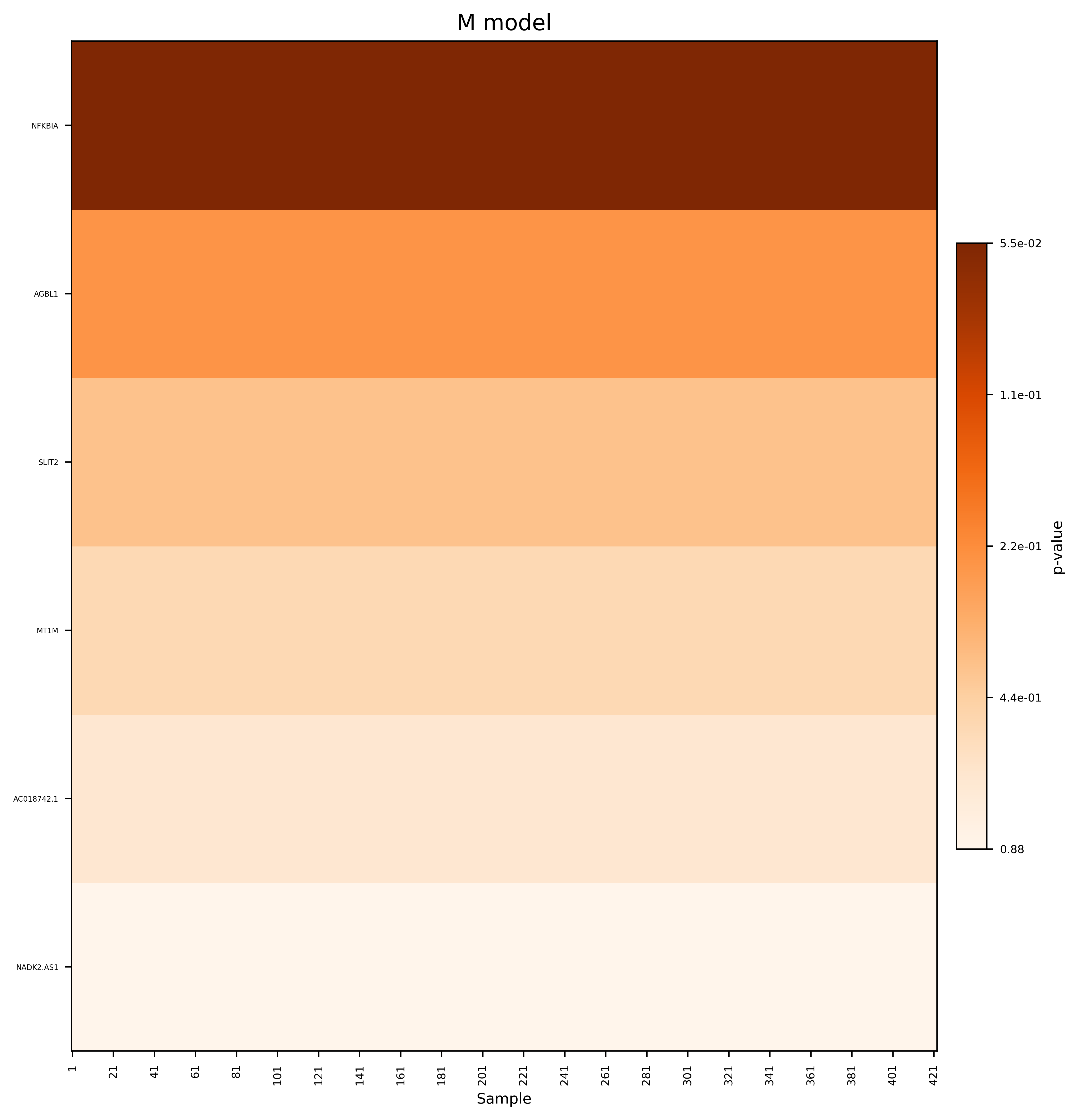}
\caption{Heatmap of p-values from the M model across samples. Each row represents a gene and each column represents a sample. Darker shades indicate smaller p-values, suggesting stronger statistical signals for mediating effects of gene-level expression levels}
\label{smore:Fig2}
\end{figure}

\begin{figure}[!ht] 
\centering
\includegraphics[scale=.44]{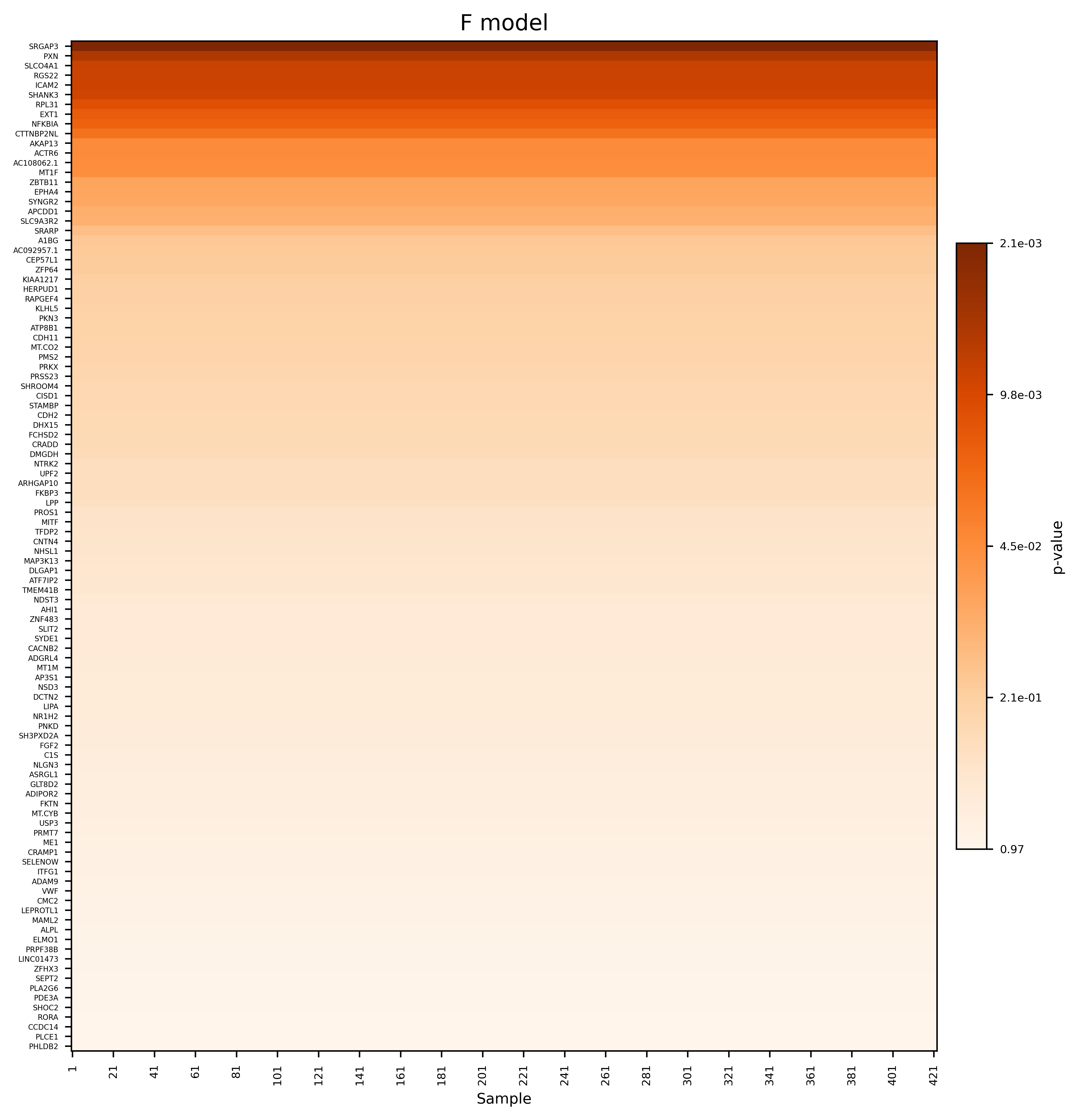}
\caption{Heatmap of p-values from the F model across samples. Each row represents a gene and each column represents a sample. Darker shades indicate smaller p-values, suggesting stronger statistical signals for mediating effects of non-expressed proportion (zero proportion)}
\label{smore:Fig3}
\end{figure}

Importantly, the greater number of genes selected by the $F$ model suggests that zero proportions may substantially influence mediation effects in the analysis of single-cell data. Among the 14 genes selected only by the $F$ model, three are highlighted here. For the SRGAP3 gene, the estimated component effects along the $X \rightarrow F$ and $F \rightarrow Y$ paths are $-14.06$ and $0.293$, respectively, resulting in a negative IIE via the $F$ model ($p_{\max} = 0.002$). Previous work has suggested that SRGAP3 plays an important role in axon guidance and neuronal connectivity, and its dysregulation has been implicated in neurodegenerative diseases including Parkinson's disease \citep{SRGAP3}. For PXN gene, the $F$ model yields a positive IIE, with estimated component effects of $-10.96$ and $-0.261$ along the $X \rightarrow F$ and $F \rightarrow Y$ paths, respectively ($p_{\max} = 0.005$). PXN has been implicated in AD across multiple omics layers. Whole-exome sequencing reported a significant accumulation of rare PXN variants in late-onset AD \citep{PXNgenetic}. Transcriptomic analyses further identified PXN as a hub gene within a dysregulated co-expression module linked to death receptor activity \citep{PXNtranscriptome}. More recently, integrative proteomic analyses also highlighted PXN in AD–associated network modules and drug-repurposing signatures \citep{PXNproteomic}. For SLCO4A1 gene, the estimated pathway effects are 9.856 and -0.255 via the $F$ model, yielding a negative indirect effect ($p_{max}$ = 0.008). A recent post-mortem cortical transcriptomic study reported significant SLCO4A1 upregulation at the mild cognitive impairment stage together with endothelial hypoxia that marks early AD-related neurovascular remodeling \citep{SLCO4A1transcriptome}. Consistently, another study identified SLO4X1 as one of the five differentially expressed oxidative-stress genes to evaluate AD subtypes and pathological outcomes in AD patients \citep{SLCO4A1GLM}. P-values from the M and F models, based on marginal modeling and regardless of statistical significance, are shown as heatmap visualizations in Figures \ref{smore:Fig2} and \ref{smore:Fig3}. The computational time for our ROSMAP data analysis was about 2 hours and 39 minutes on the high-performance Linux cluster, HiPerGator 3.0 with 15 CPU cores and 10 GB of RAM per node at University of Florida.

\begin{table*}[!htb]
\centering
\normalsize
\setlength{\tabcolsep}{11pt}
\caption{Summary of selected genes with significant mediation effects in $M$ and $F$ models (i.e., $\hat{\beta}_{M_{g}}\hat{\gamma}_{X}^{(g)} > 0$ and $\hat{\beta}_{F_{g}}\hat{\alpha}_{X}^{(g)} > 0$.)}
\label{smore:rosmaptable}
\begin{tabular}{@{} c cc cc @{}}
\toprule
 M Model & Gene & $\hat{\beta}_{M_{g}}$ & $\hat{\gamma}_{X}^{(g)}$ & $P_{\max_{g}}^{M^{1}}$  \\
\midrule 
 & NFKBIA & 4.466 & 0.234 & 0.055  \\
\midrule
 F Model & Gene & $\hat{\beta}_{F_{g}}$ & $\hat{\alpha}_{X}^{(g)}$ & $P_{\max_{g}}^{F^{2}}$  \\
 \midrule
 & AKAP13 & -4.654 & 0.354 &  0.044 \\
 & CTTNBP2NL & 6.630 & -0.308 & 0.026  \\
 & MT1F  & 10.59 & -0.297 &  0.047 \\
 & SRGAP3 & -14.06 & 0.293 &  0.002 \\
 & ICAM2 & -9.313 & -0.277 &  0.008 \\
 & NFKBIA & 12.80 & -0.283 & 0.018  \\
 & EXT1 & -5.713 & 0.244 &  0.016 \\
 & ACTR6 & 8.610 & -0.251 & 0.044 \\
 & RPL31 & -5.825 & 0.277 & 0.012  \\
 & SHANK3 & 9.163 & -0.249 &  0.008 \\
 & AC108062.1 & 9.276 & 0.256 & 0.046  \\
 & PXN & -10.96 & -0.261 & 0.005  \\
 & RGS22 & 14.45 & 0.244 & 0.008  \\
 & SLCO4A1 & 9.856 & -0.255 & 0.008  \\
\bottomrule
\end{tabular}
\begin{tablenotes}
\item [] $^{1}$ and $^{2}$ denote p-values based on the JS-test using a significance level of 0.05 or 0.1. For the F model, a threshold of 0.05 was used. No genes reached 0.05 in the M model, with the top hit below 0.1.
\end{tablenotes}
\end{table*}

\section{Discussion}
In this study, we present MedZIsc, a novel causal mediation framework tailored for single-cell data. To the best of our knowledge, this is the first methodological development in the causal mediation literature specifically designed for single-cell data. MedZIsc is built upon a BNB modeling framework, which incorporates two distinct components: (i) a beta regression model (referred to as the $F$ model throughout the paper; see Equation \ref{smore:mediationmodel1}) that captures zero-inflation by modeling the probability of structural zeros, and (ii) a NB regression model (referred to as the $M$ model throughout the paper; see Equation \ref{smore:mediationmodel2}) that accounts for overdispersed expression counts. MedZIsc is also equipped with a preliminary screening step that combines penalized regression with marginal modeling to enhance statistical power while maintaining appropriate control of the FDR. 

We defined $M_{g}$ as the average gene expression and $F_{g}$ as the non-expressed proportion across cells for each gene. This choice was motivated by an initial exploration of the scRNA-seq data, conducted prior to methodological development to better understand the distributional characteristics and sparsity nature of single-cell expression data. Specifically, we examined (i) how often a gene is expressed in all cells of a subject; (ii) how often it is not expressed in any cells of a subject; (iii) the number of genes expressed across all subjects or absent entirely; and (iv) the overall empirical distribution of gene-level zero proportions. These observations guided the formulation of the mediation framework using $M_{g}$ and $F_{g}$ as aggregated representations of expression and sparsity in this paper.

Our proposed method has practical value, as it selects mediator genes whose indirect effects (IIEs), estimated through either feature (i.e., average expression or non-expressed proportion), help explaining the association between exposure and outcome variables. This was supported and validated through our extensive simulation experiments. MedZIsc generally showed decent power in identifying true mediators, especially through the $M$ model, while maintaining a FDR control across a wide range of settings. These included from small- to large-scale simulation configurations and scenarios designed to mimic real data in terms of sample size, number of cells, and number of genes. As showcased in our real data application using the ROSMAP dataset, MedZIsc identified one gene whose average expression level and 14 genes whose non-expressed proportions in vascular or epithelial cells contributed to the mediation of the relationship between AD status and PMI with covariate adjustment. These findings are further substantiated by recent clinical studies, which have reported associations between these genes and AD, ADRD, as well as other neurodegenerative and brain disorders.

We would like to highlight several areas that present opportunities for future research. First, one possible direction is to replace the beta regression with fractional logistic regression \citep{fractionallogistic}, a method from econometrics that directly models proportion outcome bounded between 0 and 1. Second, future studies may consider extending the framework using mixed-effects models or marginal approaches, such as generalized estimating equations to account for cell-level variation. As a final remark, future research could extend the outcome model to incorporate generalized linear models (or semiparametric models), allowing the framework to handle various types of outcome variables, such as categorical outcomes (e.g., disease status or health outcome condition) or time-to-event outcomes (e.g., disease progression or patient survival).


\backmatter





\subsection*{Acknowledgments}
The authors would like to thank Mr. Mingkai Chen for his assistance in creating the heatmap figures.

\section*{Declarations}
\subsection*{Author contributions}
Conceptualization: S.A., Z.L. Methodology: S.A., Z.L. Simulation: S.A., Z.L. Software: S.A. Formal Analysis: S.A., Z.L. Writing – Original Draft: S.A. Review and Editing: L.C., M.V.G., P.R., Z.L.

\subsection*{Competing interests}
The authors declare that they have no competing interests.

\subsection*{Code availability}
The MedZIsc R package is freely available in the Comprehensive R Archive Network (CRAN) repository (\url{https://cran.r-project.org/web/packages/MedZIsc/index.html}). Please reach out to the corresponding author (Seungjun Ahn, seungjun.ahn@mountsinai.org) if you have any further inquiries.



\noindent

\bibliography{sn-bibliography}


\newpage

\newpage







\newpage 

\end{document}